\documentclass[12pt]{article}
\usepackage{amsfonts}
\usepackage{amsmath}
\usepackage{graphicx}
\usepackage{epstopdf}
\newcommand{\be}{\begin{equation}}
\newcommand{\ba}{\begin{eqnarray}}
\newcommand{\ee}{\end{equation}}
\newcommand{\ea}{\end{eqnarray}}
\newcommand{\cosech} { {\rm cosech}}
\newcommand{\sech} { {\rm sech}}

\usepackage{cite}
\usepackage{multirow}
\usepackage{setspace}
\usepackage{pdflscape} 
\usepackage{blindtext}
\begin{document}

\title{One continuous parameter family of Dirac Lorentz scalar potentials 
associated with exceptional orthogonal polynomials}

\author{Suman Banerjee$^{a}$\footnote{e-mail address: suman.raghunathpur@gmail.com(S.B)} and Rajesh Kumar Yadav$^{a}$\footnote{e-mail address: rajeshastrophysics@gmail.com(R.K.Y)}
}
 \maketitle
{$~^a$Department of Physics, Sido Kanhu Murmu University, Dumka-814110, India.\\
}

\begin{abstract}

We extend our recent works [{\it Int. J. Mod. Phys.}
A 38 (2023) 2350069-1] and obtain
one parameter $(\lambda)$ family of rationally extended Dirac Lorentz scalar 
potentials with their explicit solutions in terms of $X_{m}$ exceptional
orthogonal polynomials. We further show that as the parameter $\lambda \rightarrow 0$ or $-1$,
we get the corresponding rationally extended Pursey and the rationally extended Abraham-Moses type of scalar 
potentials respectively, which have one bound state less than the 
starting scalar potentials.

\end{abstract}

\section{Introduction}
In non-relativistic quantum mechanics, after the discovery of 
$X_{m}$-exceptional orthogonal polynomials (EOPs) \cite{dg,nr,ur},
 several new potentials have been discovered whose bound state 
solutions are in terms of these EOPs \cite{midyapd,midya1,clh11,n16,nk16,
ramos,para,nk17,nk18,rkmany,bbp,cq,op1,op2,yg15,cesp}. These potentials are generally the rational
extension of the known conventional potentials
and hence also known as rationally extended (RE) potentials. Out of these RE potentials some are exactly 
solvable and few are quasi exactly or conditionally exactly solvable potentials.
Different approaches such as 
the Darboux-Crum-transformation (DBT), Point Canonical transformation (PCT) and Supersymmetry in quantum mechanics (SQM) etc.
have been adopted to obtain these potentials and their solutions. 

One obvious question is whether there are relativistic problems whose 
solution is also in terms of these EOPs. As a first step in that direction, 
recently we \cite{srk} considered Dirac equation with Lorentz scalar potential 
and showed that indeed the exact eigenfunctions for some of the Lorentz scalar 
potentials are in terms of these EOPs. In particular, we showed that the exact
solutions for Dirac potentials (with Lorentz scalar coupling) corresponding
to RE radial oscillator potential, trigonometric scarf potential and 
generalized P\"oschl-Teller (GPT) potential can be obtained in terms of these
EOPs. One related question is, are there strictly isospectral RE Dirac 
potentials corresponding to these three Dirac potentials? Further, are 
there corresponding RE Pursey and Abraham Moses (AM) potentials with one bound 
state less? The purpose of this note is to answer these questions in the 
affirmative.
In particular, starting from these three RE Dirac scalar potentials and using
the formalism of SQM, we construct one continuous ($\lambda$) parameter family
of strictly isospectral Dirac potentials. Further, by taking the limit of 
$\lambda = 0$ and -$1$, we obtain the corresponding RE Pursey and RE 
AM potentials.

 The plan of the paper is as follows. In section $2$, we briefly discuss
the general formalism explaining how using SQM formalism one can 
construct one continuous parameter family of strictly isospectral  
Dirac Lorentz scalar potentials. We also indicate how to construct the 
corresponding Pursey and AM potentials with one bound state less.
In section 3, we apply the above formalism to the case of the RE radial 
oscillator potential and obtain the corresponding one parameter family of
strictly isospectral Dirac Lorentz scalar potentials as well as the 
corresponding RE Pursey and RE AM potentials. In order to avoid unnecessary
details, the results associated with the RE Scarf and RE GPT potentials are 
summarized in Tables $1$, $2$ and $3$. Finally, we summarize the results in 
section 4.     

\section{Formalism}

To set up the notations,  let us consider one dimensional Dirac equation with 
a Lorentz scalar potential $\phi(x)$ given by \cite{fc} 
\be\label{drc}
[i \gamma^{\mu }\partial _{\mu }- \phi(x)] \Psi_{n}(x,t)=0, \quad \mu=0,1 
\ee
where $\Psi_{n}(x,t)$ is Dirac spinor defined in the matrix form as
\be
\Psi_{n}(x,t)=
\begin{bmatrix}
\Psi^{(1)}_{n}(x,t)\\
\Psi^{(2)}_{n}(x,t)\\
\end{bmatrix}
\label{pmt}  
\ee 
If we put  $\Psi_{n}(x,t)=\exp(-i\epsilon t)\Psi_{n}(x)$ 
(where $\epsilon=$ energy associated with the spectrum) the above Dirac 
equation Eq. (\ref{drc}) takes the form
\be\label{pyt}
\gamma^{0}\epsilon \Psi_{n}(x)+i\gamma ^{1}\frac{d}{dx}\Psi_{n}(x)
-\phi_{n}(x)\Psi_{n}(x)= 0\,.
\ee
We choose a $2D$ representation of the gamma matrices \cite{jar} to directly 
cast the problem in one dimensional SUSY form i.e, 
\be
\gamma ^{0}=\sigma_{x}=
\begin{bmatrix}
0&1\\
1&0\\
\end{bmatrix}
,\qquad 
\gamma ^{1}=i\sigma_{z}=
\begin{bmatrix}
i&0\\
0&-i\\
\end{bmatrix} 
\ee\\ and
\be
\Psi_{n}(x)=
\begin{bmatrix}
\Psi ^{(1)}_{n}(x)\\
\Psi ^{(2)}_{n}(x)\\
\end{bmatrix}
\label{mno},
\ee
so that we have two coupled equations corresponding to the Dirac equation (\ref{pyt}) given by
\be\label{ijhn}
\hat{A}\Psi ^{(1)}_{n}(x)=\epsilon \Psi ^{(2)}_{n}(x) \quad \mbox{and} 
\quad \hat{A^{\dagger }}\Psi ^{(2)}_{n}(x)=\epsilon \Psi ^{(1)}_{n}(x)\,,
\ee
where the operators,  
\be\label{ghy}
\hat{A}=\frac{d}{dx}+\phi(x) \quad \mbox{and} \quad \hat{A^{\dagger }}
=-\frac{d}{dx}+\phi(x)\,.
\ee
Now the above Eq.(\ref{ijhn}) can be decoupled easily (in terms of $H_{1}(=\hat{A}\hat{A}^{\dagger})$ and\\ 
$H_{2}(=\hat{A^{\dagger}}\hat{A})$) and written as
\be
H_{2}\Psi^{(1)}_{n}(x)= \epsilon^{2}\Psi^{(1)}_{n}(x)\quad \mbox{and} 
\quad H_{1}\Psi^{(2)}_{n}(x)= \epsilon^{2}\Psi^{(2)}_{n}(x)\,,
\ee
which are equivalent to two Schr\"{o}dinger like equations namely
\be\label{ghf}
-\frac{d^{2}}{dx^{2}}\Psi ^{(1)}_{n}(x)+V^{(1)}(x)\Psi ^{(1)}_{n}(x)
=\epsilon^{2}\Psi ^{(1)}_{n}(x)\,,
\ee
\be\label{ght}
-\frac{d^{2}}{dx^{2}}\Psi ^{(2)}_{n}(x)+V^{(2)}(x)\Psi^{(2)}_{n}(x)
=\epsilon ^{2}\Psi ^{(2)}_{n}(x)\,,
\ee
with potential like terms 
\be\label{ron}
V^{(1)}(x)=\phi^{2}(x) - \phi^{'}(x)\,, 
\ee
and
\be\label{rono}
 V^{(2)}(x)=\phi^{2}(x) + \phi^{'}(x)\,.
\ee
On comparing with the well known formalism of SQM \cite{cks}, we see that
there is supersymmetry in the problem and the scalar potential $\phi(x)$ 
is just the superpotential of the SQM formalism and $V^{(1,2)}$ being the
partner potentials. Thus the eigenvalues and the eigenfunctions of the two
Hamiltonians $H_1$ and $H_2$ are related except that one of them has an extra 
bound state at zero energy so long as $\phi(x \rightarrow \pm \infty)$ have
opposite signs. Without any loss of generality we always choose $\phi(x)$
such that the ground state energy of $H_1$ is zero. In that case the 
eigenfunctions and the eigenvalues of the two Hamiltonians are related as
follows
\be\label{psi1}
\Psi_{n}^{(2)}(x) = [E_{n+1}^{(1)}]^{-1/2} A \Psi_{n+1}^{(1)}(x)\,,
\ee
\be\label{psi2}
\Psi_{n+1}^{(1)}(x) = [E_{n}^{(2)}]^{-1/2} A^{+} \Psi_{n}^{(2)}(x)\,,
\ee
\be\label{eigen}
E_{n}^{(2)} = E_{n+1}^{(1)}\,,~~E_{0}^{(1)} = 0\,,
\ee
while the scalar potential $\phi(x)$ is related to the zero energy ground state
eigenfunction by
\be\label{log}
\phi(x)=-\frac{d}{dx}[\ln\Psi_{0}^{(1)}(x)]
\ee
Finally, it is well known \cite{cks,ks,aop} that given any potential $V^{(1)}(x)$ 
with at least one bound state, it is straight forward to construct one
continuous parameter family of scalar potentials $\phi(x,\lambda)$ (which are 
nothing but the super potentials in the present case) and hence one 
continuous parameter family of potentials $V^{(1)}(x,\lambda)$. Here 
$\lambda$ can take any value $> 0$ or $<-1$. In particular,
the corresponding one parameter family of scalar potentials $\phi(x,\lambda)$
are given by \cite{ks}
\be\label{onp}
\phi(x,\lambda) = \phi(x) +\frac{d}{dx} \ln[I(x)+\lambda]\,,
\ee
where $I(x)$ in terms of the normalized ground state eigenfunction 
$\Psi_0^{(1)}(x)$ is given by
\be\label{intim}
I(x) = \int_{-\infty}^{x} [\Psi_{0}^{(1)}(y)]^2 \, dy\,.
\ee
The corresponding one continuous parameter family of strictly isospectral
potentials are
\be\label{ptnl}
V^{(1)}(x,\lambda) = V^{(1)}(x) - 2\frac{d^2}{dx^2} \ln[I(x)+\lambda]\,,
\ee
with the corresponding normalized ground state eigenfunctions being 
\be\label{zmd}
\hat{\Psi}^{(1)}_{0}(\lambda,x)=\frac{\sqrt{\lambda(1+\lambda)} 
\Psi^{(1)}_{0}(x)}{\big(I(x)+\lambda\big)}
\ee
Thus the ground state wave functions corresponding to Eq.(\ref{onp}) can be expressed as
\be
\hat{\Psi}_{0}(\lambda,x)=
\begin{bmatrix}
\hat{\Psi}^{(1)}_{0}(\lambda,x)\\
0\\
\end{bmatrix}
\label{frs1}
\ee
The normalized excited-state $(n=0,1,2....)$ eigenfunctions are obtained as-
\be
\hat{\Psi}_{n+1}(\lambda ,x)=
\begin{bmatrix}
\hat{\Psi}^{(1)}_{n+1}(\lambda ,x)\\
\Psi^{(2)}_{n}(x)\\
\end{bmatrix}
\label{frs3}\sqsupset 
\ee
where,
\be\label{extwf}
\hat{\Psi}^{(1)}_{n+1}(\lambda,x)=\Psi^{(1)}_{n+1}(x)
+\frac{1}{E^{(1)}_{n+1}}\bigg(\frac{I'(x)}{I(x)+\lambda}\bigg)
\bigg( \frac{d}{dx}+\phi(x)\bigg)\Psi^{(1)}_{n+1}(x).
\ee

Finally, as $\lambda\rightarrow 0$ and $-1$, we get the Pursey and the AM like 
scalar potentials respectively which are isospectral to $V^{(2)}(x)$. We
summarize the key features of the two potentials as below: 

(a) {\textbf{The Pursey potential:}}

  For $\lambda=0$, we get Pursey like Dirac potential from Eq.(\ref{onp}),
\be\label{spiku}
\hat{\phi}(\lambda =0,x)=\phi^{[P]}(x)=\phi(x)+\frac{d}{dx}\ln[I(x)]\,.
\ee
The corresponding eigenfunctions are
\ba\label{exrsp}
\Psi^{[P]}_{n}(x)&=&\hat{\Psi}^{(1)}_{n+1}(\lambda=0, x)\nonumber\\
&=&\Psi^{(1)}_{n+1}(x)+\frac{1}{E^{(1)}_{n+1}}\bigg(\frac{I'(x)}{I(x)}\bigg)
\bigg( \frac{d}{dx}+\phi(x)\bigg)\Psi^{(1)}_{n+1}(x),
\ea
while the energy eigen value for the $\lambda=0$ are
\be\label{skti}
E^{[P]}_n=E^{(2)}_n.
\ee 

(b) {\textbf{The Abraham-Moses potential:}}
   
   For $\lambda=-1$, we get AM like Dirac potential from Eq.(\ref{onp}),
\be\label{spjh}
\hat{\phi}(\lambda = -1,x)=\phi^{[AM]}(x)=\phi(x)+\frac{d}{dx}\ln[I(x)-1]\,.
\ee
The corresponding eigenfunctions are
\ba\label{exrsam}
\Psi^{[AM]}_{n}(x)&=&\hat{\Psi}^{(1)}_{n+1}(\lambda=-1, x)\nonumber\\
&=&\Psi^{(1)}_{n+1}(x)+\frac{1}{E^{(1)}_{n+1}}\bigg(\frac{I'(x)}{I(x)-1}\bigg)\bigg(\frac{d}{dx}+\phi(x)\bigg)\Psi^{(1)}_{n+1}(x).\nonumber\\
\ea
while the energy eigen values for $\lambda=-1$ are
\be\label{gur}
E^{[AM]}_n=E^{(2)}_n\
.\ee
Proceeding in this way, for a system with $n$ bound states, one can also obtain
n-parameter family of Lorentz scalar potential by iterating the same procedure $n$ times. 
\section{Examples}

 We now consider three different RE scalar potentials namely the RE radial 
 oscillator, RE Scarf-I and RE generalized P\"{o}schl-Teller potentials and 
 obtain one-parameter family 
of corresponding Lorentz scalar potentials with their solutions in terms of 
$X_{m}$ exceptional Orthogonal Polynomials. We discuss, the example of radial 
oscillator case in detail and summarize the results for the other two cases 
in Tables $1,2$ and $3$.

\subsection{RE Radial oscillator type Lorentz scalar potential}

In this case, we consider the $\phi(x)\rightarrow \phi_{m,ext}(r,\omega,\ell)$ given by \cite{srk}
\be\label{sonli}
\phi_{m,ext}(r,\omega,\ell) = \phi_{con}(r,\omega,\ell) + \phi_{m,rat}(r,\omega,\ell); \quad 0\leq r \leq \infty
\ee
where 
\be
\phi_{con}(r,\omega,\ell)=\frac{\omega r}{2}-\frac{(\ell+1)}{r};\quad \ell>0
\ee 
and 
\be
\phi_{m,rat}(r,\omega,\ell)=\omega r\bigg[\frac{L^{(\alpha )}_{m-1}(-z)}{L^{(\alpha-1)}_{m}(-z)}-\frac{L^{(\alpha +1)}_{m-1}(-z)}{L^{(\alpha)}_{m}(-z)}\bigg],
\ee
are the conventional and rational terms respectively. Here, 
$L^{(\alpha)}_{m}(z)$ is the Laguerre polynomial,
$z=\frac{\omega r^{2}}{2}$ and $\alpha=l+\frac{1}{2}$.
 
If we use this $\phi_{m,ext}(r,\omega,\ell)$ in equation (\ref{ron}), we get the Schr\"{o}dinger like equation (\ref{ghf}), where
\be
V^{(1)}(x)\rightarrow V^{(1)}_{m,ext}(r,\omega,\ell)= V_{con}(r,\omega,\ell) + V_{rat,m}(r,\omega,\ell)
\ee
with terms 
\be
V_{con}(r,\omega,\ell)=\frac{1}{4}\omega^2 r^2+\frac{\ell(\ell+1)}{r^2}-\omega(\ell+\frac{3}{2})
\ee
and 
\ba
V_{m,rat}(r,\omega,\ell)&=&-\omega^2r^2\frac{L^{(\alpha+1)}_{m-2}(-z)}{L_{m}^{(\alpha-1)}(-z)}+2\omega(z+\alpha-1)\frac{L^{(\alpha)}_{m-1}(-z)}{L_{m}^{(\alpha-1)}(-z)}\nonumber\\
&+&2\omega^2 r^2\bigg(\frac{L^{(\alpha )}_{m-1}(-z)}{L_{m}^{(\alpha-1)}(-z)}\bigg)^2 - 2m\omega,\quad 0<r<\infty
\ea 
which is a well-known rationally extended radial oscillator potential. The 
solution of the Eq. (\ref{ghf}) in terms of $X_{m}$-exceptional Laguerre 
polynomials $\hat{L}_{n+m}^{(\alpha)}(z)$ is thus given by
\be\label{wfrm}
\Psi^{(1)}_{n}(x)\rightarrow \Psi^{(1)}_{n,m,ext}(r,\omega,\ell)= N^{(\alpha)}_{n,m}f_{m}(\alpha,z)\hat{L}_{n+m}^{(\alpha)}(z),  
\ee
where, $f_{m}(\alpha,z)=\frac{r^{\alpha+\frac{1}{2}}
\exp\big(-\frac{z}{2}\big)}{L_{m}^{(\alpha-1)}(-z)}$
and the normalization constant 
\be
N^{(\alpha)}_{n,m}=\bigg[\frac{n!\omega^{(\alpha+1)}}{2^{\alpha }(\alpha+n+m)\Gamma(\alpha+n)}\bigg]^{1/2}
\ee 
for $m=1,2,...$ and \quad $n=0,1,2..,$. In terms of the classical Laguerre 
polynomials the expression of $\hat{L}_{m}^{(\alpha)}(z)$ is given as 
\cite{aop}
\be
\hat{L}^{(\alpha)}_{n+m}(z)=L^{(\alpha)}_m(-z)L^{(\alpha-1)}_{n}(z)+L^{(\alpha-1)}_m(-z)L^{(\alpha)}_{n-1}(z); \quad n\geq m.
\ee

If we compare Eqs. (\ref{ron}) and (\ref{rono}), we observe that the $V^{(2)}(x)\rightarrow V^{(2)}_{m,ext}(r,\omega,\ell)$ is the partner potential
of $V^{(1)}_{m,ext}(r,\omega,\ell)$ and hence the second component of the eigenfunction $\Psi_{n}(x)\rightarrow \Psi_{n,m,ext}(r,\omega,\ell)$ i.e, 
$\Psi^{(2)}_{n}(x)\rightarrow \Psi^{(2)}_{n,m,ext}(r,\omega,\ell)$
can easily be obtained using Eq. (\ref{ijhn}) i.e,
\be\label{wfprm}
\Psi^{(2)}_{n,m,ext}(r,\omega,\ell) = N^{(\alpha+1)}_{n,m}f_{m}(\alpha+1,z)\hat{L}_{n+m}^{(\alpha+1)}(z)
.\ee

The energy eigenvalues are
\be\label{evr}
E^{(1)}_{n+1}= E^{(2)}_{n} = 2(n+1) \omega\, ,~~E^{(1)}_0 = 0\,. 
\ee


Here $n=0$ corresponds to the ground state solutions.
Using the first component of the ground state Dirac eigenfunction and following Eqs.(\ref{intim}) and (\ref{onp}), 
one can obtain the integral $I(x)\rightarrow I_{m}(r,\omega,\ell)$ and hence get the one parameter isospectral family of Lorentz scalar potentials
$\hat{\phi}_{m,ext}(\lambda,r,\omega,\ell)$ for given values of $m$ and $\lambda $.

\subsubsection{Illustration for $m=1$ }

As an illustration, we consider the $X_{1}$ case $(m=1)$ for which the 
expression for the extended scalar potential (\ref{sonli}) looks like
\ba\label{evp}
\phi_{1,ext}(r,\omega,\ell)&=&\phi_{con}(r,\omega,\ell)+\phi_{1,rat}(r,\omega,\ell)\nonumber\\
&=&\frac{\omega r}{2}-\frac{(\ell+1)}{r}+\frac{4 \omega r}{(1 + 2 \ell + \omega r^2)(3 + 2\ell + \omega r^2)}\nonumber\\
\ea
The ground state Dirac eigenfunction is given by
\be\label{hjk}
\Psi^{(1)}_{0,1,ext}(r,\omega,\ell) = N^{(\alpha)}_{0,1}\Bigg(\frac{3 + 2\ell + \omega r^2}{1 + 2\ell + \omega r^2}\Bigg)r^{(1 + \ell)}\exp{(-\frac{\omega r^2}{4})}\, 
.\ee 
Now to get the explicit expression for $I_{1}(r,\omega,\ell)$, we fix the 
particular values of potential parameters (say $\omega =3$ and $\ell=1$) and 
get
 \be\label{intrad}
I_1(r,3,1)=-\frac{{e^{-\frac{{3 r^2}}{2}} \sqrt{\frac{{6}}{{\pi}}} r \left(5 + 10 r^2 + 3 r^4\right)}}
{{5 \left(1 + r^2\right)}} + \operatorname{{erf}}\left(\sqrt{\frac{{3}}{{2}}} r\right) ,
\ee
where the $\operatorname{{erf}}(r)$ is the error function.
Thus, the expression for the one parameter family (\ref{onp}) of rational 
radial oscillator type scalar potentials are given by
\ba\label{ftgd}
\hat{\phi}_{1,ext}(\lambda,r,3,1) &=&-\frac{\zeta(r) + \xi(r) (\lambda + \operatorname{{erf}}\left(\sqrt{\frac{3}{2}} r\right) )}{{2 r \left(\vartheta(r) - \Upsilon(r) (\lambda + \operatorname{{erf}}\left(\sqrt{\frac{3}{2}} r\right))\right )}},
\nonumber\\
\mbox{where,} \quad \zeta(r) &=&\sqrt{\frac{6}{\pi}} r (100 + 145 r^2 + 195 r^4 + 117 r^6 + 27 r^8),\nonumber\\
\xi(r) &=& 5 e^{\frac{3 r^2}{2}} (-20 - 9 r^2 + 12 r^4 + 9 r^6), \nonumber\\
\vartheta(r) &=& \sqrt{\frac{6}{\pi}} r (5 + 3 r^2) (5 + 10 r^2 + 3 r^4)\nonumber\\
\mbox{and}\quad \Upsilon(r) &=&5 e^{\frac{3 r^2}{2}} (5 + 8 r^2 + 3 r^4). 
\ea

The corresponding normalized ground state eigenfunction has the form
\be\label{gswf}
\hat{\Psi}^{(1)}_{0,1,ext}(\lambda,r,3,1)=\frac{\sqrt{\lambda (1+\lambda)}\Psi ^{(1)}_{0,1,ext}(r,3,1)}{[I_{1}(r,3,1)+\lambda]},
\ee
where $\Psi ^{(1)}_{0,1,ext}(r,3,1)$ and $I_{1}(r,3,1)$ 
are given by Eqs. (\ref{hjk}) and (\ref{intrad}) respectively. One can also 
obtain the normalized excited-state eigenfunctions 
\be\label{exdsr1}
\hat{\Psi}^{(1)}_{n,1,ext}(\lambda,r,3,1)=\Psi^{(1)}_{n+1}(r,3,1)+\frac{1}{E^{(1)}_{n+1}}\bigg(\frac{I_{1}'(r,3,1)}{I_{1}(r,3,1)+\lambda}\bigg)\bigg( \frac{d}{dr}+\phi_{1,ext}(r,3,1)\bigg)\Psi^{(1)}_{n+1,ext}(r,3,1)
.\ee

For $\lambda=0$, we get the RE Pursey type Lorentz scalar potentials i.e,
\ba\label{sopic}
\phi^{[P]}_{1,ext}(r)=-\frac{\zeta(r) + \xi(r) (\operatorname{{erf}}\left(\sqrt{\frac{3}{2}} r\right) )}{{2 r \left(\vartheta(r) - \Upsilon(r) (\operatorname{{erf}}\left(\sqrt{\frac{3}{2}} r\right))\right )}}
\nonumber\\
\ea
with the corresponding eigenfunctions being
\ba\label{tdrsp1}
\Psi^{[P]}_{n,1,ext}(r,3,1)=\bigg[1+\frac{1}{E^{(1)}_{n+1}}\bigg(\frac{I_{1}'(r,3,1)}{I_{1}(r,3,1)}\bigg)\bigg(\frac{d}{dr}+\phi_{1,ext}(r,3,1)\bigg)\bigg]\Psi^{(1)}_{n+1,1,ext}(r,3,1)
\ea
while using Eq.(\ref{skti}) and (\ref{evr}) the energy eigenvalues are 
given by
\be
E^{[P]}_{n}=6(n+1); 
.\ee

Similarly, for $\lambda=-1$, we get the RE AM type Lorentz scalar potentials 
\be\label{spoi1}
\phi^{[AM]}_{1,ext}(r,3,1)=-\frac{\zeta(r) + \xi(r) (-1 + \operatorname{{erf}}\left(\sqrt{\frac{3}{2}} r\right) )}{{2 r \left(\vartheta(r) - \Upsilon(r) (-1 + \operatorname{{erf}}\left(\sqrt{\frac{3}{2}} r\right))\right )}}
\nonumber\\
\ee
with the corresponding eigenfunctions being 
\ba\label{sdaram1}
\Psi^{[AM]}_{n,1,ext}(r,3,1)=\bigg[1+\frac{1}{E^{(1)}_{n+1}}\bigg(\frac{I_{1}'(r,3,1)}{I_{1}(r,3,1)-1}\bigg)\bigg(\frac{d}{dr}+\phi_{1,ext}(r,3,1)\bigg)\bigg]\Psi^{(1)}_{n+1,1,ext}(r,3,1)
.\ea

while the energy eigenvalues are 
\be
E^{[AM]}_{n}= E^{[P]}_{n} = 6(n+1).
\ee


The plots of 
$\hat{\Phi}_{1,ext}(\lambda,r,3,1)$, $\Phi^{[P]}(r,3,1)$, $\Phi^{[AM]}(r,3,1)$ 
and the normalized ground state
eigenfunctions $\hat{\Psi}^{(1)}_{0,1,ext}(\lambda,r,3,1)$ are shown in fig. $1(a)$ to $1(c)$ for different values of $\lambda$.

\includegraphics[scale=0.8]{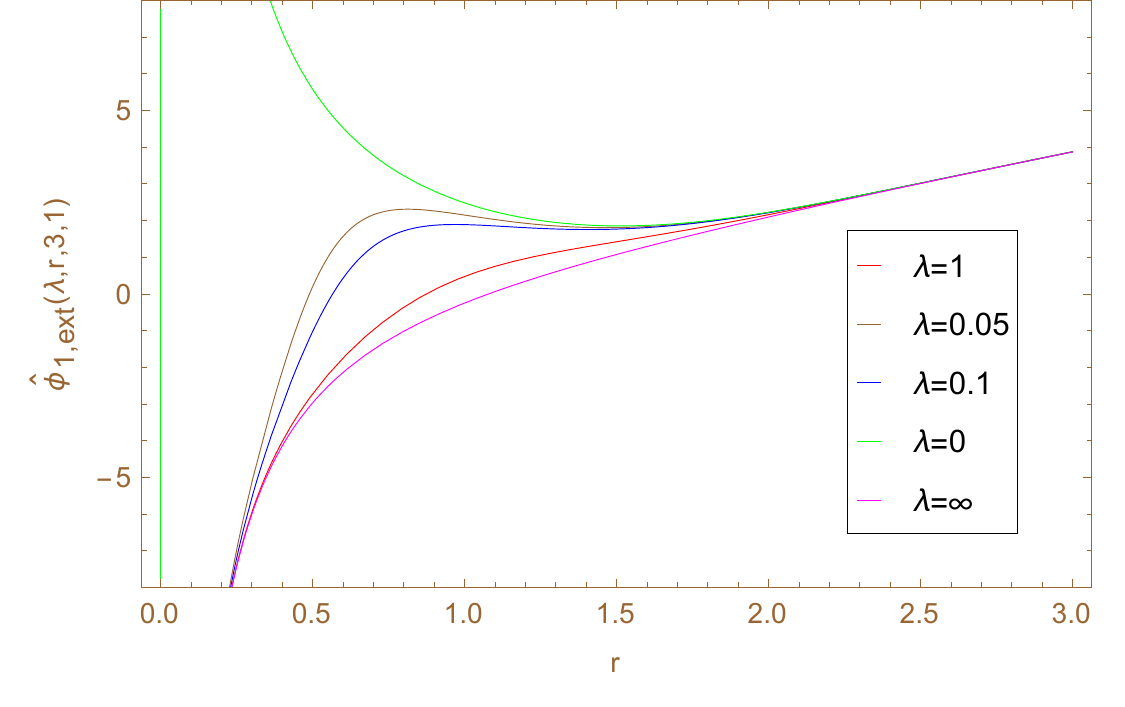}\\ 
{\bf Fig.1}: {(a) {\it Plot of $\hat{\phi}_{1,ext}(\lambda,r,3,1)$ vs. $r$ for positive $\lambda (0,0.05,0.1,1$ and $\infty)$. 
REP type scalar potential is shown for $\lambda=0$.}\\
\includegraphics[scale=0.8]{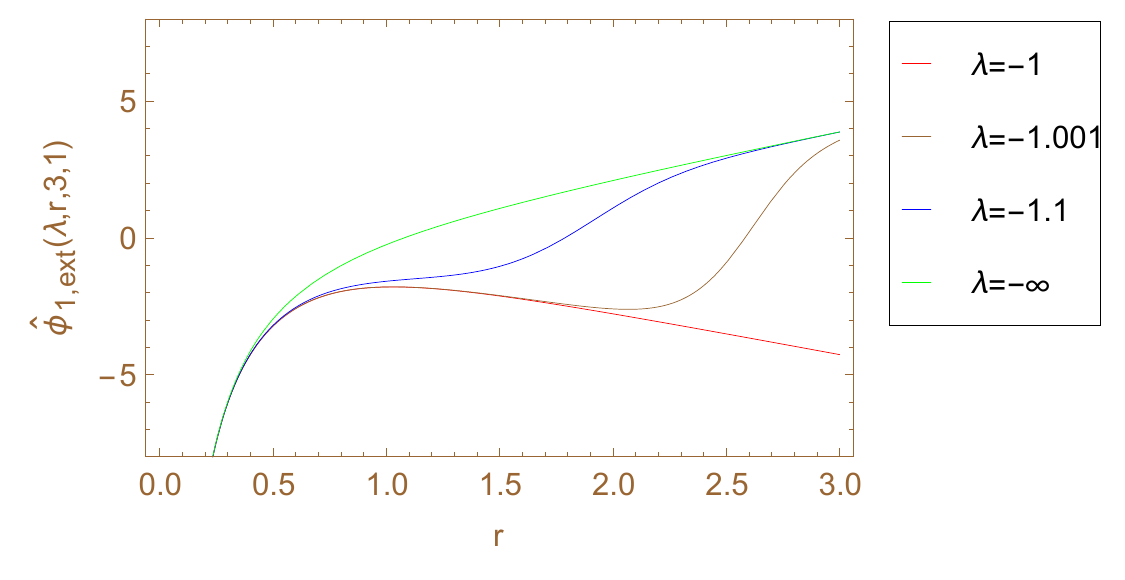}\\
{\bf Fig.1}: {(b) {\it Plots of $\hat{\phi}_{1,ext}(\lambda,r,3,1)$ 
for negative $\lambda (-\infty,-1.1,-1.01,-1.001$ and $-1)$.
The REAM type scalar potential is shown for $\lambda=-1$.}\\ 
\includegraphics[scale=0.8]{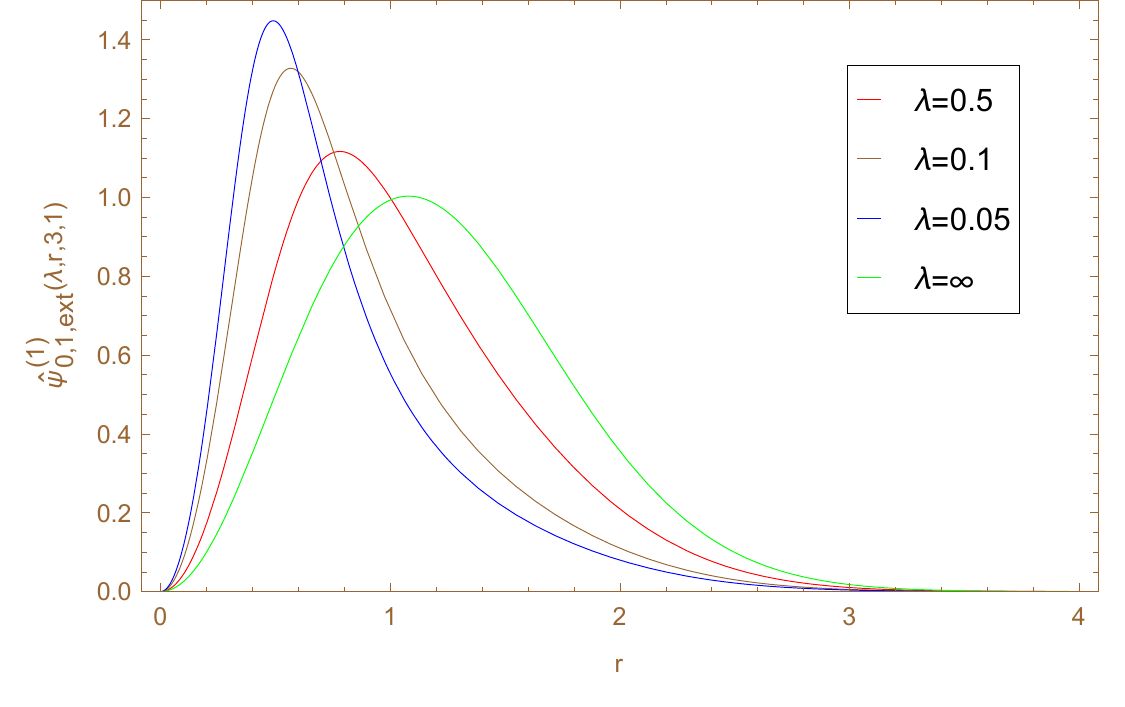}\\
{\bf Fig.1}: {(c) {\it Normalized ground-state wavefunctions $\frac{\hat{\Psi}^{(1)}_{0,1}(\lambda,r,3,1)}{r}$ for some potentials 
 (with positive $\lambda$).}\\

In Fig.1(a), we observe that as $\lambda$ starts decreasing from $+\infty$ to $0$,
the Dirac potential starts developing maxima and the corresponding peaks shifts towards $r=0$.
On the other hand in Fig.1(b), as $\lambda$ increases
from $-\infty $ to $-1$, the depth of the curves are gradually decreasing and these become
 flat as $\lambda \rightarrow -\infty$.
 
\subsection{RE Scarf-I and RE GPT type Lorentz scalar potentials}

Similar to the RE radial oscillator scalar potential case, the general form 
for the RE Scarf-I and the
RE GPT type Lorentz scalar potentials with their solutions in terms of $X_{m}$ 
Jacobi polynomials\footnote{In terms of 
classical Jacobi polynomials, the $X_{m}$-Jacobi is written as \newline
$\hat{P}_{n+m}^{(\alpha,\beta)}(z)=(-1)^m\bigg[\frac{1+\alpha+\beta+n}{2(1+\alpha+n)}(z-1)P^{(-\alpha-1,\beta-1)}_{m}(z)P^{(\alpha+2,\beta)}_{n-1}(z)
+\frac{1+\alpha-m}{\alpha+1+n}P^{(-2-\alpha,\beta)}_{m}(z)P^{(\alpha+1,\beta-1)}_{n}(z)\bigg]$.} and the corresponding 
energy eigenvalues are summarized in a tabular form and shown in Table-$1$. 
Using these we can find 
the expressions for $\phi_{m,ext}(x,A,B)$, the integral $I_{m}(x,A,B)$ and 
$\hat{\Psi}^{(1)}_{0,1,ext}(\lambda,x,A,B)$ for the special case of $m=1$ and 
fixed values of parameters $A$ and $B$ 
(for RE Scarf-I: $A=4$, $B=2$ and for RE GPT: $A=2$, $B=5$) are shown in Table:$2$. 
The one parameter family of potentials $\hat{\phi}_{1,ext}(\lambda,x,A,B)$ and 
the normalized 
ground state eigenfunctions $\hat{\Psi}^{(1)}_{0,1,ext}(\lambda,x,A,B)$ 
corresponding to these
two potentials are given in Table $3$. The expressions for the 
corresponding RE Pursey and the RE AM potentials for the same sets of
potential parameters are also given in Table $3$. The
one parameter ($\lambda $) family of rationally extended scalar potentials for 
different $\lambda$ and their corresponding ground state
eigenfunctions for RE Scarf-I and RE generalized P\"{o}schl-Teller type scalar 
Potential are shown in Fig. $2$ and Fig. $3$ respectively. 

\includegraphics[scale=0.7]{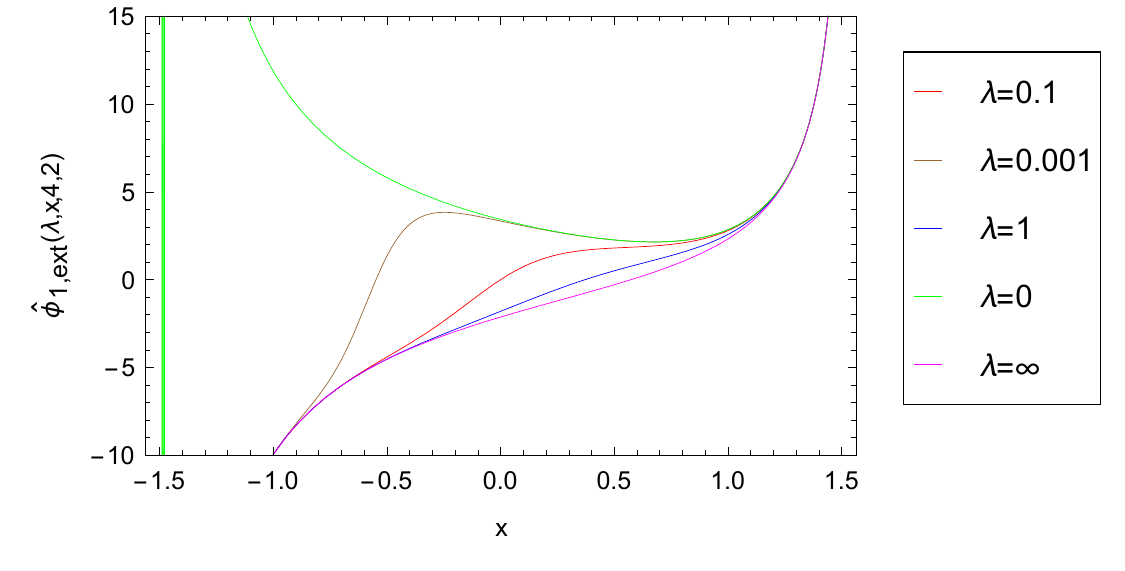}\\ 
{\bf Fig.2}: {(a) {\it Plots of $\hat{\phi}_{1,ext}(\lambda,x,4,2)$ vs. $x$ for 
positive $\lambda$ $(0,0.001,0.1,1$ and $\infty)$. The REP scalar potential is shown for $\lambda=0$.}\\
\includegraphics[scale=0.7]{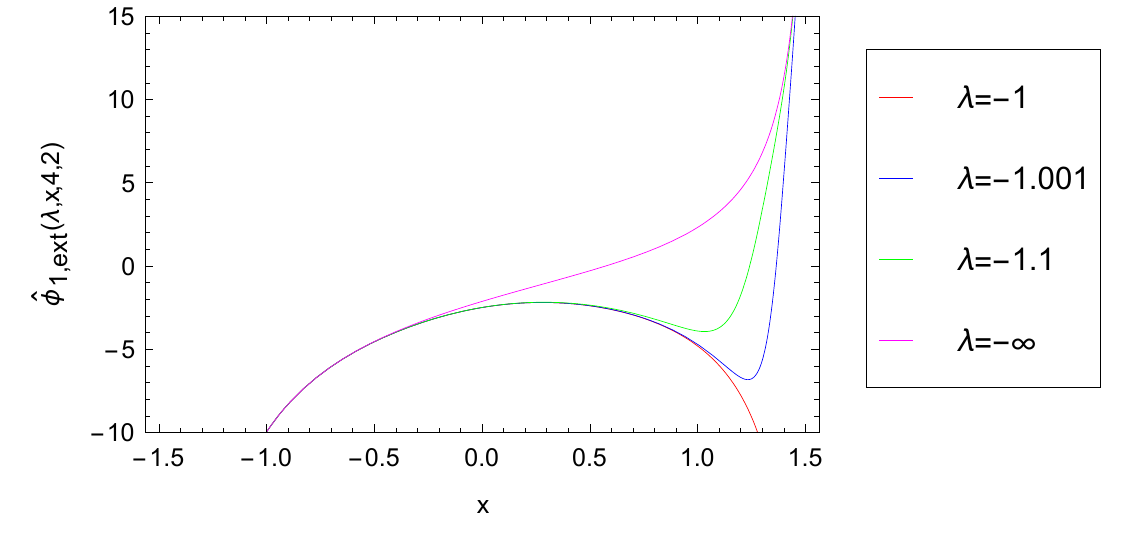} \\
{\bf Fig.2}: {(b) {\it Plots of $\hat{\phi}_{1,ext}(\lambda,x,4,2)$ vs. $r$
for negative $\lambda (-\infty,-1.001,-1.01$ and $-1)$. The REAM scalar potential is shown for $\lambda=-1$.}\\ 
\includegraphics[scale=0.7]{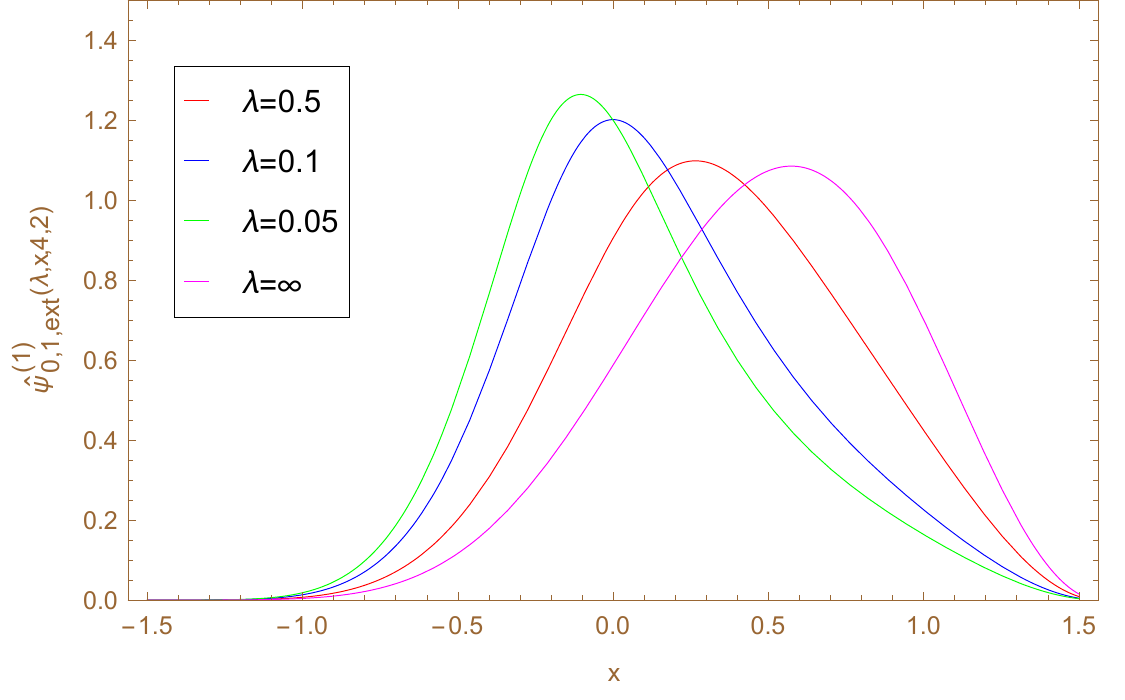}\\
{\bf Fig.2}: {(c) {\it Plots of normalized ground-state eigenfunctions $\hat{\Psi}^{(1)}_{0,1,ext}(\lambda,x,4,2)$ corresponding
one parameter family of scalar potentials (with positive $\lambda$)shown in Fig. $2(a)$. }\\

\begin{landscape}
\setlength{\arrayrulewidth}{0.5mm}
\setlength{\tabcolsep}{18pt}
\renewcommand{\arraystretch}{1.5}
\begin{table}
\begin{center}
\begin{tabular}{ |p{3cm}|p{5.8cm}|p{5.8cm}|  }
\hline
$X_{m}$ Case & RE Scarf-I & RE GPT  \\
\hline
\break $\phi _{con.}(x,A,B)$ &\break $A\tan{x} -B\sec{x}$;\newline $0<B<A-1$, \quad $-\frac{\pi }{2}< x <\frac{\pi }{2}$ \break &\break $A\coth{r} -B\cosech{r}$;\newline $B>A+1>1$, \quad $0< r <\infty$ \break \\
\hline
\break $\phi _{m,rat.}(x,A,B)$ &\break $-\frac{(\beta-\alpha+m-1)}{2}z^{'}(x) \big(\frac{P^{(-\alpha-1,\beta+1)}_{m-1}(z(x))}{P^{(-\alpha-2,\beta)}_{m}(z(x))}\ \allowbreak-\frac{P^{(-\alpha,\beta)}_{m-1}(z(x))}{P^{(-\alpha-1,\beta-1)}_{m}(z(x))}\big)$; $z(x)=\sin{x},\break$\newline $\alpha=A-B-\frac{1}{2}$,~$\beta=A+B-\frac{1}{2}$ & \break $-\frac{(\beta -\alpha + m-1)}{2}z'(r)\big(\frac{P^{(-\alpha -1,\beta +1)}_{m-1}(z(r))}{P^{(-\alpha -2,\beta )}_{m}(z(r))}-\frac{P^{(-\alpha ,\beta )}_{m-1}(z(r))}{P^{(-\alpha -1,\beta -1)}_{m}(z(r))}\big);$ $z(r)=\cosh{r}$\break \newline $\alpha=-A+B-\frac{1}{2}$,~$\beta=-A-B-\frac{1}{2}$\break \\

\hline
\break $\Psi^{(1)} _{n,m,ext}(x,A,B)$ &\break $N^{(1)}_{n,m}\frac{(1-z(x))^{\frac{(A-B)}{2}}(1+z(x))^{\frac{(A+B)}{2}}}{P^{(-\alpha-1,\beta-1)}_{m}(z(x))}\times\ \allowbreak
\hat{P}^{(\alpha,\beta)}_{n+m}(z(x))$ \break &\break $N^{(1)}_{n,m}\frac{(z-1)^{\frac{(B-A)}{2}}(z+1)^{-\frac{(B+A)}{2}}}{P^{(-\alpha-1,\beta-1)}_{m}(z(r))}\times\ \allowbreak\hat{P}^{(\alpha,\beta)}_{n+m}(z(r))$ \\
\hline
\break $E^{(1)}_{n}$ \break & \break $(A+n)^{2}-A^{2};\quad n=0,1,2..$ \break & \break $A^{2}-(A-n)^{2};\quad n=0,1,2..$ \break \\
\hline
\end{tabular}
\\[10pt]Table $1$: Expressions of Extended Lorentz scalar potentials $(\phi_{m,ext}(x,A,B)=\phi _{con.}(x,A,B)+\phi _{m,rat.}(x,A,B))$, first component of
the eigenfunctions $(\Psi^{(1)}_{n,m,ext}(x,A,B))$ and their energy eigenvalues $(E_{n}^{(1)})$ corresponding
to RE Scarf-I and RE GPT like potentials. Here $\hat{P}_{n+m}^{(\alpha,\beta)}(z)$ 
is the $X_{m}$-exceptional Jacobi orthogonal polynomial.
\end{center}
\end{table}
\end{landscape}

\begin{landscape}
\setlength{\arrayrulewidth}{0.5mm}
\setlength{\tabcolsep}{18pt}
\renewcommand{\arraystretch}{1.5}
\begin{table}
\begin{center}
\begin{tabular}{ |p{3cm}|p{6cm}|p{6cm}|  }
\hline
$X_{1} Case$ & RE Scarf-I & RE GPT  \\
\hline
\break $\phi _{1,ext}(x,A,B)$ &\break $\Bigl(4 \tan(x) - 2 \sec(x) \newline + \frac{8 \cos(x)}{-71 + 8 \cos(2x) + 64 \sin(x)}
\Bigr)$ \break & \break $\Bigl(2\coth{r} -5\cosech{r}+ 10\sinh r\times \allowbreak\big(\frac{1}{10\cosh r-5}-\frac{1}{10\cosh r-3}\big)\Bigr)$ \break \\
\hline
\break $\Psi^{(1)}_{0,1,ext}(x,A,B)$ & \break $(\frac{8}{3}\sqrt{\frac{10}{39 \pi}})\frac{(1-z(x))(1+z(x))^{3}}{P^{(-\frac{5}{2},\frac{9}{2})}_{1}(z(x))}
\hat{P}^{(\frac{3}{2},\frac{11}{2})}_{0+1}(z(x))$\break & \break \hspace{-3mm}$(21\sqrt{\frac{11}{2}})\frac{(z-1)^{\frac{(3)}{2}}(z+1)^{-\frac{(7)}{2}}}{P^{(-\frac{7}{2},-\frac{17}{2})}_{1}(z(r))}\hat{P}^{(\frac{5}{2},-\frac{15}{2})}_{0+1}(z(r))$\break \\
\hline
\vspace{15mm}$I_{1}(x,A,B)$ & \break $\Big(\frac{1}{32760 \pi (-7 + 4 \sin(x))}(-114660 (\pi + 2x)+ 244608 \cos(x)+ 59696 \cos(3x)$\newline $+ 11984 \cos(5x) - 854 \cos(7x)$\newline $- 42 \cos(9x) + 65520 (\pi + 2x)\sin(x)$\newline $- 125216 \sin(2x)+ 3416 \sin(4x)$\newline $ + 5984 \sin(6x) + 141 \sin(8x))\Big)$ \break & \break $\big(85 + 1103 \cosh(r) + 178 \cosh(2r) + 19 \cosh(3r) + \cosh(4r)\big)\times\frac{\sech^6(\frac{r}{2}) \tanh^7(\frac{r}{2})}{(-32 + 64 \cosh(r))}$ \\
\hline
\end{tabular}
\\[10pt]Table 2: The $X_{1}$ cases $(m=1)$ for RE Scarf-I and RE GPT scalar potentials, their ground state eigenfunctions
$\Psi^{(1)}_{0,1,ext}(x,A,B)$ and integral $I_{1}(x,A,B)$ for fixed values of the parameters 
$(A=4,B=2)$ and $(A=2,B=5)$ respectively.
\end{center}
\end{table}
\end{landscape}

\begin{landscape}
\setlength{\arrayrulewidth}{0.5mm}
\setlength{\tabcolsep}{18pt}
\renewcommand{\arraystretch}{1.5}
\begin{table}
\begin{center}
\scalebox{0.9}{
\begin{tabular}{ |p{1.1cm}|p{9.7cm}|p{9.7cm}|  }
\hline
$X_{1} Case$ & RE Scarf-I & RE GPT  \\
\hline
\vspace{0.005cm}\rotatebox{90}{$\hat{\phi}_{1,ext}(\lambda,x,A,B)$} &\break\hspace{-5mm}$-\frac{2 (G(x) + M(x) + 16380 (137 + 2 S(x)) (\pi + 2 x + 2 \pi \lambda)) \sec(x)}{(-9 + 4 \sin(x)) (H(x) - 114660 (\pi + 2 x + 2 \pi \lambda) + 1456 D(x) \cos(x) + 131040 \pi \lambda \sin(x))}
$\newline & $2 \coth(r) - 5 \cosech(r) + \frac{4 \sinh(r)}{3 + 4 \cosh(r)(-4 + 5 \cosh(r))}$ \newline $+\frac{198 (3 - 10 \cosh(r))^2 \cosech(r) \sinh^{7}\left(\frac{r}{2}\right)}{(-1 + 2 \cosh(r)) \left(32 \lambda \cosh^{13}\left(\frac{r}{2}\right) (-1 + 2 \cosh(r)) + Q(r) \sinh^{7}\left(\frac{r}{2}\right)\right)}$; \break \newline\hspace{-5mm} $Q(r) = 85 + 1103 \cosh(r) + 178 \cosh(2r) + 19 \cosh(3r) + \cosh(4r)$ \\
\hline
\vspace{0.005cm}\rotatebox{90}{$\hat{\Psi}^{(1)}_{0,1,ext}(\lambda,x,A,B)$} & \break $\frac{\sqrt{\lambda(1+\lambda)}}{\big(I_{1}(x,4,2)+\lambda\big)}(\frac{8}{3}\sqrt{\frac{10}{39 \pi}})\frac{(1-z(x))(1+z(x))^{3}}{P^{(-\frac{5}{2},\frac{9}{2})}_{1}(z(x))}
\hat{P}^{(\frac{3}{2},\frac{11}{2})}_{0+1}(z(x))$;\break \newline with $z(x)=\sin{x}$ & \break $\frac{\sqrt{\lambda(1+\lambda)}}{\big(I_{1}(x,2,5)+\lambda\big)}(21\sqrt{\frac{11}{2}})\frac{(z-1)^{\frac{(3)}{2}}(z+1)^{-\frac{(7)}{2}}}{P^{(-\frac{7}{2},-\frac{17}{2})}_{1}(z(r))}\hat{P}^{(\frac{5}{2},-\frac{15}{2})}_{0+1}(z(r))$;\break \newline with $z(r)=\cosh{r}$ \\
\hline
\vspace{0.005cm}\rotatebox{90}{$\phi^{[P]}_{1,ext}(x,A,B)$} &\break $-\frac{2 (G(x) + M(x) + 16380 (137 + 2 S(x)) (\pi + 2 x)) \sec(x)}{(H(x) - 114660 (\pi + 2 x) + 1456 D(x) \cos(x)) (-9 + 4 \sin(x))}
$ & \break $2 \coth\left(\frac{r}{2}\right) + \frac{10 \sinh(r)}{3 - 10 \cosh(r)} - 3 \tanh\left(\frac{r}{2}\right)$\break
\newline $+\frac{1103 \sinh(r) + 356 \sinh(2r) + 57 \sinh(3r) + 4 \sinh(4r)}{85 + 1103 \cosh(r) + 178 \cosh(2r) + 19 \cosh(3r) + \cosh(4r)}$\break\\ 
\hline
\vspace{0.5cm}\rotatebox{90}{$\phi^{[AM]}_{1,ext}(x,A,B)$} & $-\frac{2 (G(x) + M(x) - 16380 (137 + 2 S(x)) (\pi - 2 x)) \sec(x)}{(-9 + 4 \sin(x)) (H(x) + 114660 (\pi - 2 x) + 1456 D(x) \cos(x) - 131040 \pi \sin(x))}
$;\break \newline with $M(x) = -5000996 \cos(x) + 780528 \cos(3x) + 50540 \cos(5x) + 29003 \cos(7x) + 2345 \cos(9x) + 84 \cos(11x) + 3122840 \sin(2x)
$,\newline $S(x) = -35 \cos(2x) - 107 \sin(x) + 4 \sin(3x)
$, \newline $G(x) = -208840 \sin(4x) - 49335 \sin(6x) + 685 \sin(8x) + 177 \sin(10x)
$, \newline $H(x) = 59696 \cos(3x) + 11984 \cos(5x) - 854 \cos(7x) - 42 \cos(9x) + 3416 \sin(4x) + 5984 \sin(6x) + 141 \sin(8x)
$, $D(x) = 168 - 172 \sin(x) + 45 (\pi + 2 x) \tan(x)
$ & $2 \coth(r) - 5 \cosech(r) + \frac{4 \sinh(r)}{3 + 4 \cosh(r)(-4 + 5 \cosh(r))}$\break \newline $+\frac{198 (3 - 10 \cosh(r))^2 \cosech(r) \sinh^{7}\left(\frac{r}{2}\right)}{(-1 + 2 \cosh(r)) \left(-32 \cosh^{13}\left(\frac{r}{2}\right) (-1 + 2 \cosh(r)) + Q(r) \sinh^{7}\left(\frac{r}{2}\right)\right)}$; \break \newline\hspace{-5mm} $Q(r) = 85 + 1103 \cosh(r) + 178 \cosh(2r) + 19 \cosh(3r) + \cosh(4r)$ \\
\hline
\end{tabular} 
}
\\[10pt]Table $3$: The expressions for $\hat{\phi}_{1,ext}(\lambda,x,A,B)$, the
RE Pursey and the RE AM scalar potentials and their corresponding ground state
eigenfunctions ($\hat{\Psi}^{(1)}_{0,1,ext}(\lambda,x,A,B)$) for RE Scarf-I and RE GPT scalar potential for fixed values of the parameters 
$(A=4,B=2)$ and $(A=2,B=5)$ respectively.
\end{center}
\end{table}
\end{landscape}

\includegraphics[scale=0.7]{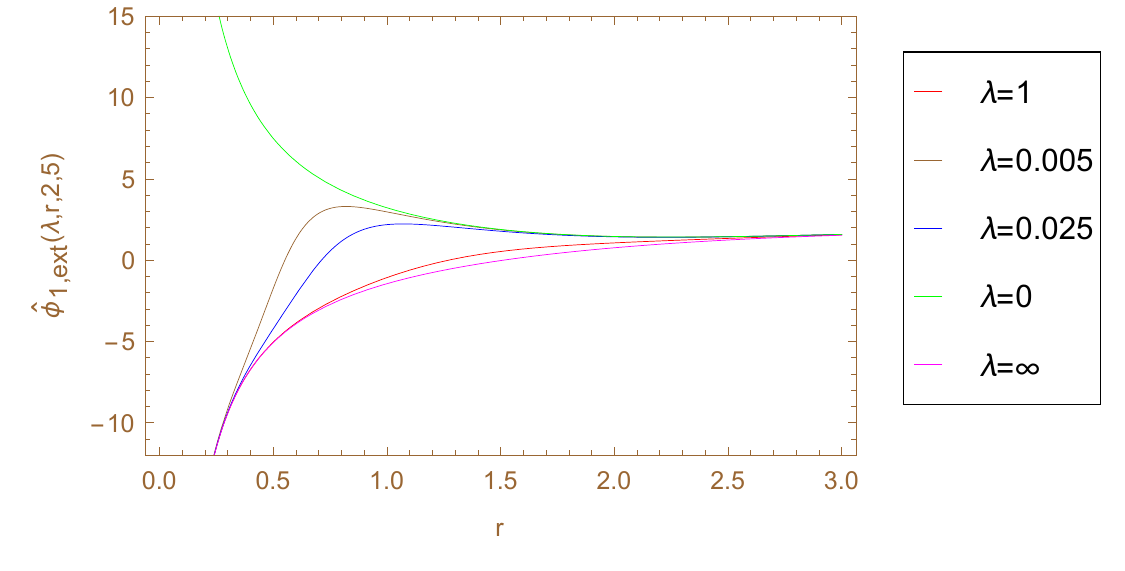}\\ 
{\bf Fig.3}: {(a) {\it Plots of $\hat{\phi}_{1,ext}(\lambda,r,2,5)$ vs. $r$ for positive 
$\lambda (0,0.001,0.01,1$ and $\infty)$. The REP potential is shown for $\lambda=0$.}\\
\includegraphics[scale=0.7]{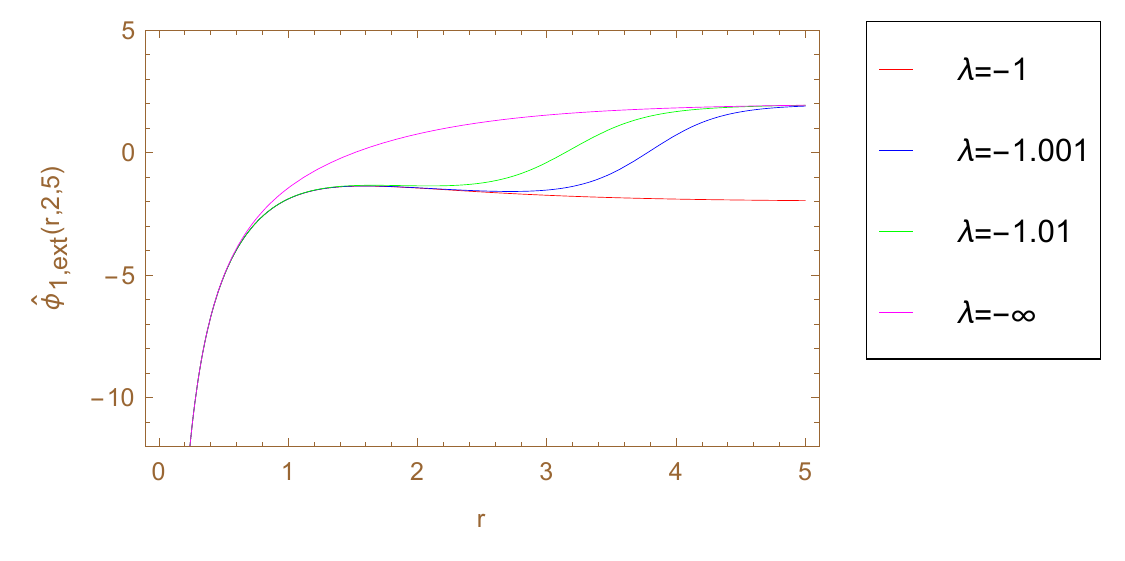} \\
{\bf Fig.3}: {(b) {\it Plots of $\hat{\phi}_{1,ext}(\lambda,r,2,5)$ vs. $r$
for negative $\lambda (-\infty,-1.1,-1.01,-1.001$ and $-1)$. The REAM scalar potential is shown for $\lambda=-1$.}\\\break 
\includegraphics[scale=0.7]{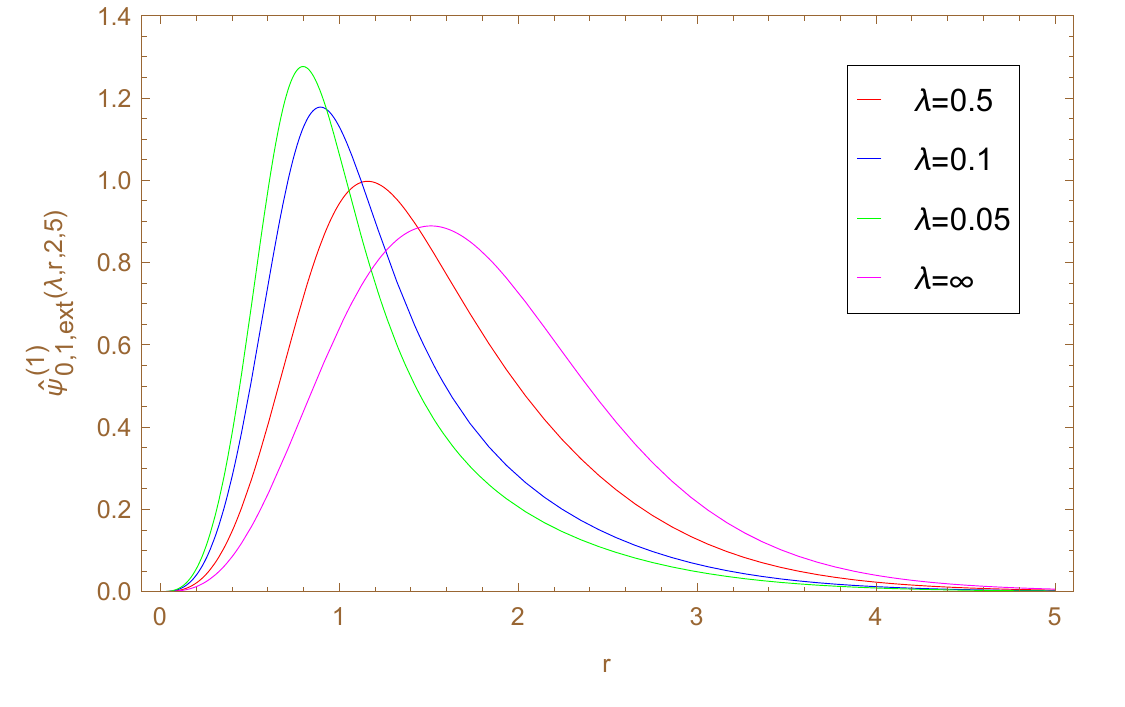}\\
{\bf Fig.3}: {(c) {\it Normalized ground-state eigenfunctions $\hat{\Psi}^{(1)}_{0,1,ext}(\lambda,r,2,5)$ for some potentials 
(with positive $\lambda$)shown in Fig. $3(a)$. }\\

\section{Summary and Conclusions}

In this work, we have considered the one-dimensional Dirac equation with three 
different RE Lorentz scalar
potentials, i.e.  
Radial oscillator, the Scarf-I and GPT scalar potentials 
and for all of them have constructed a one continuous parameter family 
$(\lambda)$ of strictly isospectral RE Lorentz
scalar potentials and obtained their solutions in terms of $X_{m}$-exceptional
orthogonal polynomials. The cases of RE Pursey and the RE AM are also discussed
in the special cases of  $\lambda =0$ and $-1$ respectively. The ground and the excited state solutions 
of all these RE strictly isospectral scalar potentials are obtained explicitly.
For the special case of $m = 1$, we have graphically shown the behavior of
the strictly isospectral potentials as a function of the continuous 
parameter $\lambda$.

\end{document}